\documentclass[a4paper,11pt,dvips]{article}
\usepackage{graphicx}
\usepackage{amsmath}
\usepackage{amssymb}
\usepackage{latexsym}
\usepackage[colorlinks]{hyperref}
\usepackage{color}
\usepackage{float}
\usepackage{cite}
\usepackage{makeidx}
\usepackage{colortbl}
\newcommand{\bra}{\begin{array}}
\newcommand{\era}{\end{array}}
\newcommand{\beq}{\begin{equation}}
\newcommand{\eeq}{\end{equation}}
\newcommand{\bqr}{\begin{eqnarray}}
\newcommand{\eqr}{\end{eqnarray}}

\def\BC{\bb C}
\def\_\BC{\bbi C}

\def\Tr {{\rm Tr}}

\def\no2 {{\textstyle{n\over 2}}}

\def\Tr {{\rm Tr}}

\newcommand{\lga}{\longrightarrow}

\begin{document}
\begin{titlepage}
\setcounter{page}{1}
\renewcommand{\thefootnote}{\fnsymbol{footnote}}

\vspace{5mm}

\begin{center}

{\Large{\bf Exact Green Function for Neutral Pauli-Dirac Particle with Anomalous Magnetic
Momentum  in Linear Magnetic Field}}

\vspace{5mm}

{\bf Abdeldjalil Merdaci\footnote{\sf amerdaci@kfu.edu.sa}}$^{a}$, {\bf Ahmed Jellal\footnote{\sf ajellal@ictp.it --
a.jellal@ucd.ac.ma}}$^{b,c}$ and {\bf Lyazid Chetouani}$^{d}$

\vspace{5mm} 
{$^{a}$ \em Physics Department, College of Science, King Faisal University,\\
PO Box
380,Alahsa 31982, Saudi Arabia}

{$^{b}$\em Theoretical Physics Group,  %Department of Physics,
Faculty of Sciences, Choua\"ib Doukkali University},\\
{\em PO Box 20, 24000 El Jadida, Morocco}

{$^c$\em Saudi Center for Theoretical Physics, Dhahran, Saudi Arabia}

{$^{d}$\em Laboratoire de Physique Math\'{e}matique et de Physique
Subatomique (LPMPS),\\ Universit\'{e} Fr\`{e}res Mentouri, 25000 Constantine, Alg\'{e}rie}

\vspace{5mm}

\begin{abstract}
We consider Pauli--Dirac
fermion submitted to an inhomogeneous magnetic field. It is showed that
the propagator of the neutral Dirac particle with an anomalous magnetic moment
in an external linear magnetic field is the causal Green function
$S^{c}(x_{b},x_{a})$ of the Pauli--Dirac equation.  The corresponding Green function
is calculated via path integral
method in global projection, giving rise to  the exact eigenspinors expressions.
The neutral particle creation probability
corresponding to our system is analyzed, which is obtained as function of the introduced field $B'$ 
and the additional spin magnetic moment $\mu$. 
\end{abstract}
\end{center}

\vspace{3cm}
\end{titlepage}

%%%%%%%%%%%%%%%%%%%%%%%%%%%%%%%%%%%%%%%%%
\section{Introduction}
%%%%%%%%%%%%%%%%%%%%%%%%%%%%%%%%%%%%%%%

In relativistic quantum mechanics, it well-know  that the Dirac equation does not correctly describe
particles of spin $1/2$. This is due to the corrections imported by the quantum electrodynamics,
such as vacuum polarization or pair creations, to  the intrinsic magnetic momentum. With this respect, the first attempt
was due to Pauli who suggested to add a term describing the anomalous of the magnetic momentum of particle
in the Dirac equation \cite{Pauli}, which is later on called  
Pauli--Dirac equation.
With this extended equation 
the exact or approximate solutions of the eigenvalue problem are very much needed.
However, it is rare to find such solutions except for interaction with 
electromagnetic fields: constant uniform magnetic field, electromagnetic plane wave and
more complicated ones \cite{Bateman1}.
In particular, 
we recall that 
the Dirac equation for charged and neutral fermions with anomalous magnetic moments was solved
in a uniform magnetic field \cite{Pitschmanna}. The expressions of relativistic wave functions and energy spectra were explicitly obtained. 
It was  showed 
that, in the non–relativistic limit the wave functions and energy spectra of charged fermions agreed with the known
solutions of the Schr\"odinger equation.
The dependence of the
relativistic wave functions on the magnetic quantum number was discussed in more detail
and the obtained results  were compared with the literature \cite{Seetharaman, Chand}.

In theoretical physics there is a formalism that became a powerful mathematical tool, %for theoretician that
this is the path integral introduced by Feynman years ago. Even it is successfully applied  in the field theory and quantum mechanics
but  faced serious problems with the relativistic quantum mechanics. %Specifically, 
Such problems are related to
the covariance that requires treating the space and time in equally as wall as taking into account of
the dynamics of spin as an additional degree of freedom. In 1951, the first problem was solved by  
Schwinger \cite{Schwinger},
who used 
an approach based on the proper time parametrization and allowed him
to explicitly determine the corrected magnetic momentum. However,
the second problem is much more complicated because of the nature of the spin momentum. More precisely, there is an interplay 
between two sets such that the spin is a 
discrete object (quantum mechanics) and the path integral
is based on the trajectory notion (classical mechanics). 
To overcome such difficulties and 
after several attempts to construct the Green function corresponding to the relativistic
particles using the path integral mechanism, Fradkin and Gitman were succeeded in establishing an interesting mathematical formalism
\cite{Fradkin}.

Motivated by different works and specifically reference \cite{Pitschmanna}, we study 
the problem of the relativistic
Pauli-Dirac neutral particle submitted to a linear magnetic field by adopting the path integral mechanism. In the first step, 
we show 
that the global
projection gives rise the Green function of the inverse Pauli--Dirac
operator.
This will be used together with some algebra to explicitly
determine the corresponding eigenspinors as well as analyze their asymptotic behaviors.
Subsequently, we determine  the probability of pair creation $\cal{W}$ in vacuum in 2+1 dimensions corresponding 
to our system. This can be done by making use the effective action together with
the charge conjugation matrix to end up with $\cal{W}$ in terms of the additional spin magnetic momenta and 
magnetic field.

The paper is organized as follows. In section 2, we show that how one can use
the global projection for (2 + 1)-dimensional Pauli-Dirac equation in
linear magnetic field. In section 3, we give the theoretical formulation of
the problem where different changes are introduced to simplify the process for
obtaining the solutions of eigenspinors. More precisely, we will use the
causal Green function as well as different techniques to solve our problem. We
determine the eigenspinors in terms of different physical parameters and
quantum numbers in section 4. In section 5, we calculate the probability of
pair creation in terms of different parameters. We conclude our
findings in the final section.

%%%%%%%%%%%%%%%%%%%%%%%%%%%%%%%%%%%%%%%%%%
\section{Global projection}
%%%%%%%%%%%%%%%%%%%%%%%%%%%%%%%%%%%%%%%%%%%

%It is known that
%Recalling that
We start by recalling that 
the Clifford algebra with three generating elements has two
inequivalent two-dimensional irreducible representations of gamma matrices.
These will serve as a guide %allow us %to achieve our gaol by describing
to describe the behavior of a system of relativistic charged particle in constant magnetic field
with a given value of the helicity. 
In doing so, we show 
that the global
projection giving rise the Green function of the inverse Pauli--Dirac
operator $O_{-}^{-1}$
\begin{equation}
S^{c}=\left(  \gamma_{\varsigma}^{\mu}p_{\mu}-m-\frac{\mu}{2}\sigma^{\mu\nu
}F_{\mu\nu}\right)  ^{-1}=O_{-}^{-1}=O_{+}\left(  O_{-}O_{+}\right)  ^{-1}
\label{eq1}%
\end{equation}
where the operators $O_{\pm}$ read as
\begin{equation}
O_{\pm}=\gamma_{\varsigma}^{\mu}p_{\mu}-\frac{\mu}{2}\sigma^{\mu\nu}F_{\mu\nu
}\pm m \label{eq2}%
\end{equation}
while $\mu$ describes the additional spin magnetic moment. The matrices
$\gamma_{\varsigma}^{\mu}$ are defined through the following relations 
\begin{equation}
\left[  \gamma_{\varsigma}^{\mu},\gamma_{\varsigma}^{\nu}\right]  _{+}%
=2\eta^{\mu\nu},\qquad\left[  \gamma_{\varsigma}^{\mu},\gamma_{\varsigma}%
^{\nu}\right]  _{-}=-2i\sigma^{\mu\nu} \label{eq3}%
\end{equation}
generating the Clifford
algebra, which is giving rise to the algebra of different
representations for Dirac gamma matrices, labeled by the subscript
$\varsigma=\pm1$. particularly
in (2+1)-dimensions  the metric is
%Actually in odd dimensions, we have 
$\eta^{\mu\nu}=\mbox{diag}\left(
1,-1,-1\right)$ and the matrices are mapped as
\beq
\gamma_{\varsigma}^{\mu}%
=\frac{i\varsigma}{2}\epsilon^{\mu\nu\lambda}\gamma_{\varsigma,\nu}%
\gamma_{\varsigma,\lambda}, \qquad  \mu,\nu=0,1,2
\eeq
or equivalently
\begin{equation}
\gamma_{\varsigma}^{0}=i\varsigma\gamma_{\varsigma}^{1}\gamma_{\varsigma}%
^{2}=\sigma_{3},\qquad\gamma_{\varsigma}^{1}=-i\varsigma\gamma_{\varsigma}%
^{0}\gamma_{\varsigma}^{2}=i\sigma_{2},\qquad\gamma_{\varsigma}^{2}%
=-i\varsigma\gamma_{\varsigma}^{0}\gamma_{\varsigma}^{1}=-i\varsigma\sigma
_{1}. \label{eq4}%
\end{equation}
 Now from \eqref{eq1}, one can see that the Green function $S^{c} $
satisfies the equation%
\begin{equation}
\left(  \gamma_{\varsigma}^{\mu}p_{\mu}-m-\frac{\mu}{2}\sigma^{\mu\nu}%
F_{\mu\nu}\right)  S^{c}=-\mathbb{I}_{3\otimes3}. \label{eq5}%
\end{equation}
Note that, formally $S^{c}(\mathbf{x}_{b},\mathbf{x}_{a})$ is the matrix element in the
coordinate space and therefore it can be written as
\begin{equation}
S^{c}(\mathbf{x}_{b},\mathbf{x}_{a})=\langle\mathbf{x}_{b}|S^{c}%
|\mathbf{x}_{a}\rangle.\label{eq6}%
\end{equation}
It is clearly seen that, if we multiply
%Now multiplying 
both sides of \eqref{eq5} by the states $\left\langle \mathbf{x}_{b}\right\vert$ and
$\left\vert \mathbf{x}_{a}\right\rangle$,  we end up with  the following equation
for $S^{c}(\mathbf{x}_{b},\mathbf{x}_{a})$ 
%satisfying the Dirac--Pauli equation
%\begin{subequations}
\begin{equation}
\left(  \gamma_{\varsigma}^{\mu}\left(  i\partial_{b\mu}\right)  -m-i\frac
{\mu}{2}\gamma_{\varsigma}^{\mu}\gamma_{\varsigma}^{\nu}F_{\mu\nu}(
\mathbf{x}_{b})S^{c}(\mathbf{x}_{b},\mathbf{x}_{a})\right)
%=-\left\langle \mathbf{x}_{b}\mid\mathbf{x}_{a}\right\rangle 
=-\delta
^{3}(\mathbf{x}_{b}-\mathbf{x}_{a}) \label{eq7}%
\end{equation}
which is nothing but the Pauli--Dirac equation that will play a crucial role
in our analysis and allows to deal with our task.

Having the expressions for the operators $O_{+}$ and $O_{-}$, we can define the Green operator for global projection by
%\end{subequations}
\begin{equation}
S_{g}^{c}=\left(  O_{-}O_{+}\right)  ^{-1}=\left(  O_{+}O_{-}\right)  ^{-1}
\label{eq8}%
\end{equation}
where the label $g$ stands for global. One can easily show that %It is clear that 
the matrix element of
$S_{g}^{c}$ verifies the quadratic Dirac equation
\begin{equation}
O_{-}O_{+}S_{g}^{c}(\mathbf{x}_{b},\mathbf{x}_{a})=O_{+}O_{-}S_{g}%
^{c}(\mathbf{x}_{b},\mathbf{x}_{a})=-\delta^{3}(\mathbf{x}_{b}-\mathbf{x}_{a})
\label{eq9}%
\end{equation}
Now we can establish a relation between 
the matrix elements $S^{c}(\mathbf{x}_{b},\mathbf{x}_{a})$
and $S^{c}_g(\mathbf{x}_{b},\mathbf{x}_{a})$ of both projections.
Indeed, 
using \eqref{eq2}, \eqref{eq7} and \eqref{eq9} to end up with 
%the relation
%between $S^c$ and $S^c_g$
%Therefore, we can deduce the following form
\begin{equation}
S^{c}(\mathbf{x}_{b},\mathbf{x}_{a})=\left(  \gamma^{\mu}\left(
i\partial_{b\mu}\right)  +m-i\frac{\mu}{2}\gamma_{\varsigma}^{\mu}%
\gamma_{\varsigma}^{\nu}F_{\mu\nu}\left(  \mathbf{x}_{b}\right)  \right)
S_{g}^{c}(\mathbf{x}_{b},\mathbf{x}_{a}). \label{eq10}%
\end{equation}
In the next, we show how the above mathematical tools can be used to deal with different
issues those 
concern 
the eigenspinors of our system as well as the probability of pair creation in vacuum.

%%%%%%%%%%%%%%%%%%%%%%%%%%%%%%%%%%%%%%%%%%%%%%%%%%%%%%%%%%%%%%
%\section{Green function calculations in global projection}
%%%%%%%%%%%%%%%%%%%%%%%%%%%%%%%%%%%%%%%%%%%%%%%%%%%%%%%%%%%%
%%%%%%%%%%%%%%%%%%%%%%%%%%%%%%%%%%%%%%%%%%%%%%%%%%%%%%%%%%%%%%
\section{Green function} % and global projection}
%%%%%%%%%%%%%%%%%%%%%%%%%%%%%%%%%%%%%%%%%%%%%%%%%%%%%%%%%%%%

To achieve our goals, 
%To apply the above mathematical tools, 
let us fix our system by
%In this section we 
considering the interaction of a neutral fermion with a linear
magnetic field $B$. %More precisely, our system is composed of $\left(
%2+1\right)$-dimensional Dirac neutral fermions in an external magnetic field. 
It is 
convenient for our task  to choose an interaction described by the following 
quadri-potential
\begin{equation}
A_{0}=0,\qquad A_{x}=0,\qquad A_{y}=Bx+\frac{1}{2}B^{\prime}x^{2} \label{eq11}
\end{equation}
where the field $B'$ is introduced to generalize the constant magnetic field cases
and more importantly will allow to deal with the pair creation. % of two different magnetic fields.
%In this case,
One can use this gauge to show that 
the operators \eqref{eq2} take the forms
\begin{equation}
O_{\pm}=\sigma_{3}\left(  \hat{p}_{0}-\varsigma\mu\frac{B+B^{\prime}\hat{x}%
}{2}\right)  -i\sigma_{2}\hat{p}_{x}+i\varsigma\sigma_{1}\hat{p}_{y}\pm m.
\label{eq12}%
\end{equation}
These allow us to define %Furthermore, 
$S_{g}%
^{c}(\mathbf{x}_{b},\mathbf{x}_{a})$ as %may also be defined as
\begin{eqnarray}
S_{g}^{c}(\mathbf{x}_{b},\mathbf{x}_{a}) &=&\langle\mathbf{x}_{b}\mid\left(
O_{+}O_{-}\right)  ^{-1}\mid\mathbf{x}_{a}\rangle\\
&=&-\frac{i}{2}\int_{0}%
^{\infty}d\lambda\ \langle\mathbf{x}_{b}\mid\exp\left(  i\frac{\lambda}%
{2}\left(  \mathcal{H}+i\varepsilon\right)  \right)  \mid\mathbf{x}_{a}%
\rangle\label{eq13}\nonumber
\end{eqnarray}
%With the Feynman $i\varepsilon$ prescription. 
where $i\varepsilon$ is the Feynman  prescription and
the involved Hamiltonian $\mathcal{H}=O_{+}O_{-}=O_{-}O_{+}$  is given by
\begin{equation}
\mathcal{H}    =\left(  \hat{p}_{0}-\varsigma\mu
\frac{B+B^{\prime}x}{2}\right)  ^{2}-\hat{p}_{x}^{2}-\hat{p}_{y}^{2}%
-m^{2}+i\varsigma\frac{\mu B^{\prime}}{2}\sigma_{1}
\end{equation}
which can be written under the unitary transformation $e^{+i\frac{\pi}{4}\sigma_{2}}$
as
\begin{equation}
 \mathcal{H}  
=e^{-i\frac{\pi}{4}\sigma_{2}}\left[  \left(  \hat{p}_{0}-\varsigma\mu
\frac{B+B^{\prime}x}{2}\right)  ^{2}-\hat{p}_{x}^{2}-\hat{p}_{y}^{2}%
-m^{2}+i\varsigma\frac{\mu B^{\prime}}{2}\sigma_{3}\right]  e^{+i\frac{\pi}%
{4}\sigma_{2}} \label{eq14}.%
\end{equation}

In the path integral representation, 
%We present 
$S_{g}^{c}\left(  \mathbf{x}_{b},\mathbf{x}_{a}\right)$ 
takes the form
%by means
%of a path integral%
\begin{eqnarray}
S_{g}^{c}\left(  \mathbf{x}_{b},\mathbf{x}_{a}\right)   &  =&-\frac{i}{2}%
\int_{0}^{\infty}d\lambda\int DtDxDy\int Dp_{0}Dp_{x}Dp_{y}\ e^{-i\frac{\pi}%
{4}\sigma_{2}}e^{-\varsigma\frac{\mu B^{\prime}}{4}\sigma_{3}}e^{+i\frac{\pi
}{4}\sigma_{2}}\\
&& \exp\left(  i\int_{0}^{1}\left[  p_{0}\dot{t}-p_{x}\dot{x}-p_{y}%
\dot{y}+\frac{\lambda}{2}\left(  \left(  p_{0}-\varsigma\mu\frac{B+B^{\prime
}x}{2}\right)  ^{2}-p_{x}^{2}-p_{y}^{2}-m^{2}\right)  \right]  d\tau\right)
\nonumber\label{eq15}%
\end{eqnarray}
where $\mathbf{x=}\left(  t,x,y\right)  $ satisfying the boundary conditions
\begin{equation}
\mathbf{x}(0)=\mathbf{x}_{a},\qquad\mathbf{x}(1)=\mathbf{x}_{b} \label{eq16}.
\end{equation}
Integrating over the paths $t$ and $y$, one can see that the momenta become
constants, i.e. $p_{0}$=const and $p_{y}$=const. Again integrating once more over
$p_{x}$, to rewrite $S_{g}^{c}\left(  \mathbf{x}_{b},\mathbf{x}_{a}\right)  $
as
\begin{eqnarray}
S_{g}^{c}\left(  \mathbf{x}_{b},\mathbf{x}_{a}\right)   &  =& -\frac{i}{2}%
\int_{0}^{\infty}d\lambda\int_{-\infty}^{+\infty}\frac{dp_{0}}{2\pi}%
\frac{dp_{y}}{2\pi}\ e^{-ip_{0}\left(  t_{b}%
-t_{a}\right)  +ip_{y}\left(  y_{b}-y_{a}\right)  -\frac{i\lambda}{2}\left(
p_{y}^{2}+m^{2}\right)  }\nonumber\\
& &  e^{-i\frac{\pi}{4}\sigma_{2}}e^{-\lambda\varsigma\frac{\mu
B^{\prime}}{4}\sigma_{3}}e^{+i\frac{\pi}{4}\sigma_{2}}K^{os}\left(
x_{a},x_{b};\lambda\right)  \label{eq17}%
\end{eqnarray}
where $K^{os}$ is the propagator related to the  paths $x(\tau)$ and expresses
the motion of a particle submitted to an harmonic oscillator with an imaginary
frequency $\omega=i\tfrac{\mu B^{\prime}}{2}$ and mass $m=1$, which is
\begin{equation}
K^{os}\left(  x_{a},x_{b};\lambda\right)  =\int
Dx\ e^{i\int_{0}^{1}\left(  \frac{\dot{x}^{2}}{2\lambda}+\lambda\frac{\mu
^{2}B^{\prime2}}{8}\left(  x-\frac{2\varsigma p_{0}-\mu B}{\mu B^{\prime}%
}\right)  ^{2}\right)  d\tau}. \label{eq18}%
\end{equation}
The integral over the paths $x(\tau)$ is well-known and equal to%
\begin{equation}
K^{os}\left(  x_{a},x_{b};\lambda\right)  =\left(  \tfrac{\tfrac{\mu
B^{\prime}}{2}}{2i\pi\sinh\left(  \tfrac{\lambda\mu B^{\prime}}{2}\right)
}\right)  ^{1/2}\exp\left(  i\tfrac{\left(  z_{b}^{2}+z_{a}^{2}\right)
\cosh\left(  \tfrac{\lambda\mu B^{\prime}}{2}\right)  -2z_{b}z_{a}}%
{4\sinh\left(  \tfrac{\lambda\mu B^{\prime}}{2}\right)  }\right)  \label{eq19}%
\end{equation}
where $z=\sqrt{\mu B^{\prime}}\left(  x-\frac{2\varsigma p_{0}-\mu B}{\mu
B^{\prime}}\right)$. The propagator $K^{os}$ can be expressed as
\cite{barut}%
\begin{equation}
K^{os}    =\frac{1}{\sqrt{1+\xi^{2}}}\exp\left[  \frac{1}{4}\frac{1-\xi^{2}%
}{1+\xi^{2}}\left(  \alpha^{2}+\beta^{2}\right)  +i\alpha\beta\frac{\xi}%
{1+\xi^{2}}\right]
\end{equation}
or  in terms of the parabolic cylindrical function  $D_{\nu}\left(  z\right)  $ 
\begin{equation}
K^{os}    =\frac{\left(  \tfrac{\pi}{2}\right)  ^{1/2}}{2\pi i}\int_{\iota-i\infty}^{\iota+i\infty}\frac
{\xi^{-\left(  \nu+1\right)  }}{\sin\left(
-\pi\nu\right)  }
{\sum_{\epsilon=\pm1}}
%EndExpansion
D_{\nu}\left(  \epsilon e^{-i\frac{\pi}{4}}\alpha\right)  D_{-\nu-1}\left(
\epsilon e^{i\frac{\pi}{4}}\beta\right)  d\nu\label{eq20}%
\end{equation}
and the used notations are $-1<\iota<0,
\xi=e^{i\frac{\pi}{2}}e^{\tfrac{\lambda\mu B^{\prime}}{2}} (\left\vert \arg\xi\right\vert <\frac{\pi}{2}), \alpha
=z_{b},\beta=z_{a}$.
After
inserting \eqref{eq20} into \eqref{eq17} we obtain%
\begin{eqnarray}
S_{g}^{c}\left(  \mathbf{x}_{b},\mathbf{x}_{a}\right)   &  =&-%\frac{1}{2}%
\frac{1}{4\pi }\left(
\tfrac{\mu B^{\prime}}{4}\right)  ^{1/2}\int_{0}^{\infty}d\lambda\int
_{-\infty}^{+\infty}\frac{dp_{0}}{2\pi}\frac{dp_{y}%
}{2\pi}e^{-ip_{0}\left(  t_{b}-t_{a}\right)  +ip_{y}\left(  y_{b}%
-y_{a}\right)  -\frac{i\lambda}{2}\left(  p_{y}^{2}+m^{2}\right)  }\nonumber\\
&&  \int_{\iota-i\infty}^{\iota+i\infty}\tfrac{\xi^{-\left(  \nu+1/2\right)  }}{\sin\left(  -\pi\nu\right)
}e^{-i\frac{\pi}{4}\sigma_{2}}e^{-\varsigma\frac{\mu B^{\prime}}{4}\sigma_{3}} e^{i\frac{\pi}{4}\sigma_{2}}%
{\sum_{\epsilon=\pm1}}
%EndExpansion
D_{\nu}\left(  \epsilon e^{-i\frac{\pi}{4}}z_{b}\right)  D_{-\nu-1}\left(
\epsilon e^{i\frac{\pi}{4}}z_{a}\right)  d\nu\label{eq21}.
\end{eqnarray}
%where we have used%
%\begin{equation}
%-1<\iota<0,\qquad\left\vert \arg\xi\right\vert <\frac{\pi}{2},\qquad
%\xi=e^{i\frac{\pi}{2}}e^{\tfrac{\lambda\mu B^{\prime}}{2}},\qquad\alpha
%=z_{b},\beta=z_{a} \label{eq22}%
%\end{equation}

%At this stage, 
To go further in developing \eqref{eq21},
let us introduce the spin operator $\sigma_{3}$ and insert the identity
$\sum_{s=\pm1}\chi_{s}\chi_{s}^{+}=\mathbb{I}_{2\otimes2}$. We can easily check the
relations
\begin{equation}
\sigma_{3}\chi_{s}=s\chi_{s},\qquad\sigma_{1}\chi_{s}=\chi_{-s},\qquad
\sigma_{2}\chi_{s}=is\chi_{-s}\label{eq23}%
\end{equation}
for the vectors
\begin{equation}
\chi_{+1}=\left(
\begin{array}
[c]{c}%
1\\
0
\end{array}
\right)  ,\qquad\chi_{-1}=\left(
\begin{array}
[c]{c}%
0\\
1
\end{array}
\right)\label{eq24}%
\end{equation}
which 
can be implemented in \eqref{eq21} to obtain the causal Green function $S_{g}^{c}\left(  \mathbf{x}_{b},\mathbf{x}_{a}\right)$
\begin{eqnarray}
S_{g}^{c} &  =&-\frac{\left(  \mu B^{\prime}\right)^{1/2} }%
{8\pi} \int_{0}^{\infty}d\lambda
\int_{-\infty}^{+\infty}\frac{dp_{0}}{2\pi}
\frac{dp_{y}}{2\pi}e^{-ip_{0}\left(  t_{b}-t_{a}\right)  +ip_{y}\left(
y_{b}-y_{a}\right)  -\frac{i\lambda}{2}\left(  p_{y}^{2}+m^{2}\right)
}
\sum_{s=\pm1} e^{-\varsigma\frac{\mu B^{\prime}}{4}s} \nonumber\\
&& 
e^{-i\frac
{\pi}{4}\sigma_{2}}\chi_{s}\chi_{s}^{+}e^{+i\frac{\pi}{4}\sigma_{2}}
\int_{\iota-i\infty}^{\iota+i\infty}\tfrac{\xi^{-\left(  \nu+1/2\right)  }}{\sin\left(  -\pi\nu\right)
}e^{-i\frac{\pi}{4}\sigma_{2}}%
%TCIMACRO{\tsum _{s=\pm1}}%
%BeginExpansion
%{\sum_{s=\pm1}}
%EndExpansion%
%TCIMACRO{\tsum _{\epsilon=\pm1}}%
%BeginExpansion
{\sum_{\epsilon=\pm1}}
%EndExpansion
D_{\nu}\left(  \epsilon e^{-i\frac{\pi}{4}}z_{b}\right)  D_{-\nu-1}\left(
\epsilon e^{i\frac{\pi}{4}}z_{a}\right)  d\nu.\label{eq25}%
\end{eqnarray}
%As $\varepsilon=\pm1$ and $\varsigma s=\pm1$, one can easily check the
Using the identity%
\begin{equation}
\sum_{s=\pm1}f_{s}\sum_{\varepsilon=\pm1}g_{\varepsilon}=\sum_{s=\pm1}%
f_{s}\left(  g_{s}+g_{-s}\right)\label{eq26}%
\end{equation}
to write 
$S_{g}^{c}\left(  \mathbf{x}_{b},\mathbf{x}_{a}\right)  $ as
\begin{eqnarray}
S_{g}^{c} &  =& -\frac{\left(  \mu B^{\prime}\right)  ^{1/2}}{8\pi}
\int_{0}^{\infty}d\lambda\\
&&\int_{-\infty}^{+\infty}\frac{dp_{0}}{2\pi
} \frac{dp_{y}}{2\pi}
e^{-ip_{0}\left(  t_{b}%
-t_{a}\right)  +ip_{y}\left(  y_{b}-y_{a}\right)  -\frac{i\lambda}{2}\left(
p_{y}^{2}+m^{2}\right)  } \sum_{s=\pm1} e^{-\varsigma\frac{\mu B^{\prime}}{4}s}e^{-i\frac
{\pi}{4}\sigma_{2}}\chi_{s}\chi_{s}^{+}e^{+i\frac{\pi}{4}\sigma_{2}%
}\nonumber\\
&& \int_{\iota-i\infty}^{\iota+i\infty}\tfrac{\xi^{-\left(  \nu+1/2\right)  }}{\sin\left(  -\pi\nu\right)  }%
%TCIMACRO{\tsum _{s=\pm1}}%
%BeginExpansion
%{\sum_{s=\pm1}}
%EndExpansion
\left[  D_{\nu}\left(  se^{-i\frac{\pi}{4}}z_{b}\right)  D_{-\nu-1}\left(
se^{i\frac{\pi}{4}}z_{a}\right)  +D_{\nu}\left(  -se^{-i\frac{\pi}{4}}%
z_{b}\right)  D_{-\nu-1}\left(  -se^{i\frac{\pi}{4}}z_{a}\right)  \right]
d\nu. \nonumber\label{eq27}
\end{eqnarray}
Once we integrate over $\lambda,$ we obtain  $\left[  \frac
{i}{2}\left(  p_{y}^{2}+m^{2}-i\mu B^{\prime}\left(  \nu+\frac{1+\varsigma
s}{2}\right)  \right)  \right]  ^{-1}$, which appears at the denominator and
out from the following pole%
\begin{eqnarray}
\nu %&=&-\tfrac{1+\varsigma s}{2}-i\tfrac{p_{y}^{2}+m^{2}}{\mu B^{\prime}}%
%\nonumber\\
=-\tfrac{1+\varsigma s}{2}-i\rho\label{eq28}%
\end{eqnarray}
where we have introduced the quantity $\rho=\tfrac{p_{y}^{2}+m^{2}}{\mu
B^{\prime}}$.
We finally obtain
%\begin{eqnarray} \label{eq100}
%S_{g}^{c} &  =&-\tfrac{4\pi i}{\mu B^{\prime}}\int_{-\infty}^{+\infty}%
%\frac{dp_{0}}{2\pi} %\int_{-\infty}^{+\infty}
%\frac{dp_{y}}{2\pi}e^{-ip_{0}%
%\left(  t_{b}-t_{a}\right)  +ip_{y}\left(  y_{b}-y_{a}\right)  }
%\sum_{s=\pm1}
%EndExpansion
%\frac{\frac{e^{i\frac{\pi}{4}\varsigma s-\frac{\pi}{2}\rho}}{8\pi}\left(  \mu
%B^{\prime}\right)  ^{1/2}\varsigma s}{-i\sinh\left(  \pi\rho\right)
%}e^{-i\frac{\pi}{4}\sigma_{2}}\chi_{s}\chi_{s}^{+}e^{+i\frac{\pi}{4}\sigma
%_{2}}\\
%&&\left[  D_{-\frac{1+\varsigma s}{2}-i\rho}\left(  se^{-i\frac{\pi}%
%{4}}z_{b}\right)  D_{-\frac{1-\varsigma s}{2}+i\rho}\left(  se^{i\frac{\pi}%
%{4}}z_{a}\right)  +D_{-\frac{1+\varsigma s}{2}-i\rho}\left(  -se^{-i\frac{\pi
%}{4}}z_{b}\right)  D_{-\frac{1-\varsigma s}{2}+i\rho}\left(  -se^{i\frac{\pi
%}{4}}z_{a}\right)  \right]\nonumber . %
%\end{eqnarray}
\begin{eqnarray}\label{eq100}
 S_{g}^{c} &  =& \int_{-\infty}^{+\infty}\frac{dp_{0}}{2\pi} %\int_{-\infty}^{+\infty}%
\frac{dp_{y}}{2\pi}e^{-ip_{0}\left(  t_{b}-t_{a}\right)  +ip_{y}\left(
y_{b}-y_{a}\right)  }%
%TCIMACRO{\tsum _{s=\pm1}}%
%BeginExpansion
{\sum_{s=\pm1}}
%EndExpansion
\frac{e^{i\pi\frac{\varsigma s}{4}-\frac{\pi}{2}\rho}\varsigma s}{2\left(  \mu
B^{\prime}\right)  ^{1/2}\sinh\left(  \pi\rho\right)  }e^{-i\frac{\pi}%
{4}\sigma_{2}}\chi_{s}\chi_{s}^{+}e^{+i\frac{\pi}{4}\sigma_{2}}\\
&&\left[  D_{-\frac{1+\varsigma s}{2}-i\rho}\left(  se^{-i\frac{\pi}%
{4}}z_{b}\right)  D_{-\frac{1-\varsigma s}{2}+i\rho}\left(  se^{i\frac{\pi}%
{4}}z_{a}\right)  +D_{-\frac{1+\varsigma s}{2}-i\rho}\left(  -se^{-i\frac{\pi
}{4}}z_{b}\right)  D_{-\frac{1-\varsigma s}{2}+i\rho}\left(  -se^{i\frac{\pi
}{4}}z_{a}\right)  \right]. \nonumber
\end{eqnarray}
For more information about %It is interesting to note that 
the parabolic cylindrical functions of the
form  $D_{-\tfrac{1}{2}+i\alpha}\left(  \pm e^{-i\frac{\pi}{4}}z\right)
$ and their properties, we refer to \cite{Abramowitz}. 
%\textbf{are presented with their properties in more detailed in} \cite{Shinnosuke}.

%%%%%%%%%%%%%%%%%%%%%%%%%%%%%%%
\section{Eigenspinors}
%%%%%%%%%%%%%%%%%%%%%%%%%%%%%%%%%

At this stage, we
show how the obtained results above can be used to %in order to
explicitly 
%We would like to 
determine the eigenspinors corresponding to our system. To this end, we
map \eqref{eq100} into 
%consider 
the causal Green function \eqref{eq10} 
%together with \eqref{eq100}
to end up with 
\begin{eqnarray}
S^{c}  &  =&\left[  \left(  i\tfrac{\partial}{\partial t_{b}}-\varsigma\mu
\frac{B+B^{\prime}x_{b}}{2}\right)  \sigma_{3}-i\sigma_{2}\left(
-i\tfrac{\partial}{\partial x_{b}}\right)  +i\varsigma\sigma_{1}\left(
-i\tfrac{\partial}{\partial y_{b}}\right)  +m\right] \nonumber \\
&&  \int_{-\infty}^{+\infty}\frac{dp_{0}}{2\pi} %\int_{-\infty}^{+\infty}%
\frac{dp_{y}}{2\pi}e^{-ip_{0}\left(  t_{b}-t_{a}\right)  +ip_{y}\left(
y_{b}-y_{a}\right)  }%
%TCIMACRO{\tsum _{s=\pm1}}%
%BeginExpansion
{\sum_{s=\pm1}}
%EndExpansion
\frac{e^{i\pi\frac{\varsigma s}{4}-\frac{\pi}{2}\rho}\varsigma s}{2\left(  \mu
B^{\prime}\right)  ^{1/2}\sinh\left(  \pi\rho\right)  }e^{-i\frac{\pi}%
{4}\sigma_{2}}\chi_{s}\chi_{s}^{+}e^{+i\frac{\pi}{4}\sigma_{2}}\\
&&\left[  D_{-\frac{1+\varsigma s}{2}-i\rho}\left(  se^{-i\frac{\pi}%
{4}}z_{b}\right)  D_{-\frac{1-\varsigma s}{2}+i\rho}\left(  se^{i\frac{\pi}%
{4}}z_{a}\right)  +D_{-\frac{1+\varsigma s}{2}-i\rho}\left(  -se^{-i\frac{\pi
}{4}}z_{b}\right)  D_{-\frac{1-\varsigma s}{2}+i\rho}\left(  -se^{i\frac{\pi
}{4}}z_{a}\right)  \right] \nonumber \label{label eq30}%
\end{eqnarray}
which gives 
the Green function relative to our particle as
\begin{eqnarray}\label{label eq31}%
S^{c}(\mathbf{x}_{b},\mathbf{x}_{a})  &  =& \int_{-\infty}^{+\infty}\frac
{dp_{0}}{2\pi} %\int_{-\infty}^{+\infty}
\frac{dp_{y}}{2\pi}e^{-ip_{0}\left(
t_{b}-t_{a}\right)  +ip_{y}\left(  y_{b}-y_{a}\right)  }%
%TCIMACRO{\tsum _{s=\pm1}}%
%BeginExpansion
{\sum_{s=\pm1}}
%EndExpansion
\frac{e^{i\pi\frac{\varsigma s}{4}-\frac{\pi}{2}\rho}\varsigma s}{2
\sinh\left(  \pi\rho\right)  } e^{-i\frac{\pi}{4}\sigma_{2}}\\
&& \left[  \left(  -is\sqrt{\mu B^{\prime
}}\left(  \tfrac{\partial}{\partial z_{b}}+i\tfrac{s\varsigma}{2}z_{b}\right)
\chi_{-s}\chi_{s}^{+}+\left(  i\varsigma sp_{y}+m\right)  \chi_{s}\chi_{s}%
^{+}\right)  \right]  e^{+i\frac{\pi}{4}\sigma_{2}} 
\left[  D_{-\frac{1+\varsigma s}{2}-i\rho}\left(  se^{-i\frac{\pi}%
{4}}z_{b}\right) \right. \nonumber\\
&& \left. D_{-\frac{1-\varsigma s}{2}+i\rho}\left(  se^{i\frac{\pi}%
{4}}z_{a}\right)  +D_{-\frac{1+\varsigma s}{2}-i\rho}\left(  -se^{-i\frac{\pi
}{4}}z_{b}\right)  D_{-\frac{1-\varsigma s}{2}+i\rho}\left(  -se^{i\frac{\pi
}{4}}z_{a}\right)  \right].  \nonumber
\end{eqnarray}
For later use we introduce
%Now by applying 
the following recurrence and derivative relations between the parabolic
cylindrical functions\cite{Bateman}
\begin{eqnarray}
 && zD_{\nu}\left(  z\right)     =D_{\nu+1}\left(  z\right)  +\nu D_{\nu
-1}\left(  z\right)\\
&& \tfrac{\partial}{\partial z}D_{\nu}\left(  z\right)   =\frac{1}{2}\nu
D_{\nu-1}\left(  z\right)  -\frac{1}{2}D_{\nu+1}\left(  z\right)
\label{label eq32}.
 % \label{label eq33}%
\end{eqnarray}
After verification for all values $\varsigma s=\pm1$, we obtain  the
interesting property%
\begin{equation}
\left(  \frac{d}{dz}-\frac{s\varsigma}{2}z\right)  D_{-\frac{1+\varsigma s}%
{2}-i\rho}\left(  z\right)  =-\tfrac{\left(  1-s\varsigma\right)  \rho+\left(
1+s\varsigma\right)  }{2}D_{-\frac{1-\varsigma s}{2}-i\rho}\left(  z\right)
\label{label eq34}%
\end{equation}
which implies the relation
%This can be used to easily
%show easily the relation
\begin{equation}
\left(  \frac{d}{dz_{b}}-\frac{s\varsigma}{2}z_{b}\right)  D_{-\frac
{1+\varsigma s}{2}-i\rho}\left(  se^{-i\frac{\pi}{4}}z_{b}\right)
=-se^{-i\frac{\pi}{4}s\varsigma}\tfrac{\left(  1-s\varsigma\right)
\rho+\left(  1+s\varsigma\right)  }{2}D_{-\frac{1-\varsigma s}{2}-i\rho
}\left(  se^{-i\frac{\pi}{4}}z_{b}\right). \label{label eq35}%
\end{equation}
Using all to %This can be used to
write $S^{c}(\mathbf{x}_{b},\mathbf{x}_{a})$ as%
\begin{eqnarray}\label{eq36}
S^{c}(\mathbf{x}_{b},\mathbf{x}_{a})  &  = &\int_{-\infty}^{+\infty}\frac
{dp_{0}}{2\pi} %\int_{-\infty}^{+\infty}
\frac{dp_{y}}{2\pi}e^{-ip_{0}\left(
t_{b}-t_{a}\right)  +ip_{y}\left(  y_{b}-y_{a}\right)  }%
%TCIMACRO{\tsum _{s=\pm1}}%
%BeginExpansion
{\sum_{s=\pm1}}
%EndExpansion
\frac{e^{-\frac{\pi}{2}\rho}\varsigma}{2\sinh\left(  \pi\rho\right)
}\nonumber\\
&& e^{-i\frac{\pi}{4}\sigma_{2}}\left\{  is\tfrac{\left(
1-s\varsigma\right)  \rho+\left(  1+s\varsigma\right)  }{2}D_{-\frac
{1-\varsigma s}{2}-i\rho}\left(  se^{-i\frac{\pi}{4}}z_{b}\right)
D_{-\frac{1-\varsigma s}{2}+i\rho}\left(  se^{i\frac{\pi}{4}}z_{a}\right)
\chi_{-s}\chi_{s}^{+}\right. \nonumber\\
&&  -is\tfrac{\left(  1-s\varsigma\right)  \rho+\left(  1+s\varsigma\right)
}{2}D_{-\frac{1-\varsigma s}{2}-i\rho}\left(  -se^{-i\frac{\pi}{4}}%
z_{b}\right)  D_{-\frac{1-\varsigma s}{2}+i\rho}\left(  -se^{i\frac{\pi}{4}%
}z_{a}\right)  \chi_{-s}\chi_{s}^{+}\\
&&  +se^{i\frac{\pi}{4}s\varsigma}\tfrac{i\varsigma sp_{y}+m}{\sqrt{\mu
B^{\prime}}}D_{-\frac{1+\varsigma s}{2}-i\rho}\left(  se^{-i\frac{\pi}{4}%
}z_{b}\right)  D_{-\frac{1-\varsigma s}{2}+i\rho}\left(  se^{i\frac{\pi}{4}%
}z_{a}\right)  \chi_{s}\chi_{s}^{+}\nonumber\\
&&  \left.  +se^{i\frac{\pi}{4}s\varsigma}\tfrac{i\varsigma sp_{y}+m}{\sqrt{\mu
B^{\prime}}}D_{-\frac{1+\varsigma s}{2}-i\rho}\left(  -se^{-i\frac{\pi}{4}%
}z_{b}\right)  D_{-\frac{1-\varsigma s}{2}+i\rho}\left(  -se^{i\frac{\pi}{4}%
}z_{a}\right)  \chi_{s}\chi_{s}^{+}\right\}  e^{+i\frac{\pi}{4}\sigma_{2}}.\nonumber
%\label{label eq36}%
\end{eqnarray}
In developing further the above relation, let us introduce
%By  implementing 
the mapping
$s\longrightarrow-s$ in \eqref{eq36} but only for the terms $D_{-\frac{1+\varsigma s}{2}-i\rho}\left(
-se^{-i\frac{\pi}{4}}z_{b}\right)  D_{-\frac{1-\varsigma s}{2}+i\rho}\left(
-se^{i\frac{\pi}{4}}z_{a}\right)$, to get the form
%in  the Green function take
%the following form%
\begin{eqnarray}
S^{c}(\mathbf{x}_{b},\mathbf{x}_{a})  &  =&\int_{-\infty}^{+\infty}\frac
{dp_{0}}{2\pi} %\int_{-\infty}^{+\infty}
\frac{dp_{y}}{2\pi}e^{-ip_{0}\left(
t_{b}-t_{a}\right)  +ip_{y}\left(  y_{b}-y_{a}\right)  }%
%TCIMACRO{\tsum _{s=\pm1}}%
%BeginExpansion
{\sum_{s=\pm1}}
%EndExpansion
\frac{e^{-\frac{\pi}{2}\rho}\varsigma}{2\sinh\left(  \pi\rho\right)
}\nonumber\\
&&   e^{-i\frac{\pi}{4}\sigma_{2}}is\left\{  \tfrac{\left(
1-s\varsigma\right)  \rho+\left(  1+s\varsigma\right)  }{2}D_{-\frac
{1-\varsigma s}{2}-i\rho}\left(  sZ_{b}\right)
D_{-\frac{1-\varsigma s}{2}+i\rho}\left(  sZ_{a}\right)
\chi_{-s}\chi_{s}^{+}\right. \nonumber\\
&&  +\tfrac{\left(  1+s\varsigma\right)  \rho+\left(  1-s\varsigma\right)  }%
{2}D_{-\frac{1+\varsigma s}{2}-i\rho}\left(  sZ_{b}\right)
D_{-\frac{1+\varsigma s}{2}+i\rho}\left(  sZ_{a}\right)
\chi_{s}\chi_{-s}^{+}\\
&&  -ie^{+i\frac{\pi}{4}s\varsigma}\tfrac{i\varsigma sp_{y}+m}{\sqrt{\mu
B^{\prime}}}D_{-\frac{1+\varsigma s}{2}-i\rho}\left(  sZ_{b}\right)  D_{-\frac{1-\varsigma s}{2}+i\rho}\left(  sZ_{a}\right)  \chi_{s}\chi_{s}^{+}\nonumber\\
&&  \left.  +ie^{-i\frac{\pi}{4}s\varsigma}\tfrac{-i\varsigma sp_{y}+m}%
{\sqrt{\mu B^{\prime}}}D_{-\frac{1-\varsigma s}{2}-i\rho}\left(
sZ_{b}\right)  D_{-\frac{1+\varsigma s}{2}+i\rho}\left(
sZ_{a}\right)  \chi_{-s}\chi_{-s}^{+}\right\}
e^{+i\frac{\pi}{4}\sigma_{2}} \label{label eq37} \nonumber%
\end{eqnarray}
where we have used the notations $Z_a= e^{i\frac{\pi}{4}}z_{a}$ and
$Z_b=e^{-i\frac{\pi}{4}}z_{b}$.
By factorizing we obtain
\begin{eqnarray}
&&S^{c}(\mathbf{x}_{b},\mathbf{x}_{a})  = i\varsigma%
%TCIMACRO{\tsum _{s=\pm1}}%
%BeginExpansion
{\sum_{s=\pm1}}
%EndExpansion
s\int_{-\infty}^{+\infty}\frac{dp_{0}}{2\pi} %\int_{-\infty}^{+\infty}%
\frac{dp_{y}}{2\pi}e^{-ip_{0}\left(  t_{b}-t_{a}\right)  +ip_{y}\left(
y_{b}-y_{a}\right)  }
%\nonumber\\
%&&  %
%TCIMACRO{\tsum _{s=\pm1}}%
%BeginExpansion
%{\sum_{s=\pm1}}
%EndExpansion
\tfrac{e^{-\frac{\pi}{2}\rho}}{2\sinh\left(  \pi\rho\right)  }e^{-i\frac{\pi
}{4}\sigma_{2}}
\left\{  \left[  \tfrac{\left(  1+s\varsigma\right)
\frac{i\varsigma sp_{y}+m}{\sqrt{\mu B^{\prime}}}+\left(  1-s\varsigma\right)
}{2} \right. \right. \nonumber\\
&& \left. \left. D_{-\frac{1+\varsigma s}{2}-i\rho}\left(  sZ_{b}\right)  \chi_{s}\right.  \right. 
+\left.  i\tfrac{\left(  1-s\varsigma\right)  \frac{-i\varsigma sp_{y}%
+m}{\sqrt{\mu B^{\prime}}}+\left(  1+s\varsigma\right)  }{2}e^{-i\frac{\pi}%
{4}s\varsigma}D_{-\frac{1-\varsigma s}{2}-i\rho}\left(  sZ_{b}\right)  \chi_{-s}\right] 
%\nonumber\\
%&&
\left[  \tfrac{\left(  1+s\varsigma\right)  \frac{-i\varsigma
sp_{y}+m}{\sqrt{\mu B^{\prime}}}+\left(  1-s\varsigma\right)  }{2}%
\right. \nonumber\\
&& \left.  \left. D_{-\frac{1+\varsigma s}{2}+i\rho}\left(  sZ_{a}\right)
\chi_{s}^{+}
  -  i\tfrac{\left(  1-s\varsigma\right)  \frac{i\varsigma
sp_{y}+m}{\sqrt{\mu B^{\prime}}}+\left(  1+s\varsigma\right)  }{2}%
e^{+i\frac{\pi}{4}s\varsigma}D_{-\frac{1-\varsigma s}{2}+i\rho}\left(
sZ_{a}\right)  \chi_{-s}^{+}\right]  \right\}  \sigma
_{1}e^{+i\frac{\pi}{4}\sigma_{2}} \label{label eq38}%
\end{eqnarray}
which can be further  simplified to end up with
\begin{equation}
S_{\varsigma}^{c}(\mathbf{x}_{b},\mathbf{x}_{a})=i\varsigma\sum_{s=\pm1}%
s\int_{-\infty}^{+\infty}\frac{dp_{0}}{2\pi} %\int_{-\infty}^{+\infty}%
\frac{dp_{y}}{2\pi}\Psi_{p_{0}p_{y}}^{\left(  \varsigma,s\right)  }\left(
x_{b},y_{b};t_{b}\right)  \bar{\Psi}_{p_{0}p_{y}}^{\left(  \varsigma
,s\right)  }\left(  x_{a},y_{a};t_{a}\right)   \label{label eq39}%
\end{equation}
with $\bar{\Psi}=\Psi^{+}\sigma_{3}$ and the normalized eigenspinors are given in
compact form as
\begin{equation}
\Psi_{p_{0}p_{y}}^{\left(  \varsigma,s\right)  }\left(  x,y;t\right)
=u_{p_{0}p_{y}}^{\left(  \varsigma,s\right)  }\left(  x,y;t\right)  \chi
_{s}+v_{p_{0}p_{y}}^{\left(  \varsigma,s\right)  }\left(  x,y;t\right)
\chi_{-s} \label{label eq40}%
\end{equation}
where $u_{n,s}$ and $v_{n,s}$ are two-component defined by%
\begin{eqnarray}
u_{p_{0},p_{y}}^{\left(  \varsigma,s\right)  }\left(  x,y;t\right)   &
=& \frac{1}{2}\tfrac{e^{-\frac{\pi}{4}\rho}}{\sqrt{\sinh\left(  \pi\frac
{p_{y}^{2}+m^{2}}{\mu B^{\prime}}\right)  }}e^{-ip_{0}t}e^{+ip_{y}%
y}\nonumber\\
&&  \left\{  \tfrac{\left(  1+s\varsigma\right)  \frac{i\varsigma sp_{y}%
+m}{\sqrt{\mu B^{\prime}}}+\left(  1-s\varsigma\right)  }{2}D_{-\frac
{1+\varsigma s}{2}-i\frac{p_{y}^{2}+m^{2}}{\mu B^{\prime}}}\left(
se^{-i\frac{\pi}{4}}\sqrt{\mu B^{\prime}}\left(  x-\tfrac{2\varsigma p_{0}-\mu
B}{\mu B^{\prime}}\right)  \right)  \right. \\
&&  \left.  -is\tfrac{\left(  1-s\varsigma\right)  \frac{-i\varsigma sp_{y}%
+m}{\sqrt{\mu B^{\prime}}}+\left(  1+s\varsigma\right)  }{2}e^{-i\frac{\pi}%
{4}s\varsigma}D_{-\frac{1-\varsigma s}{2}-i\frac{p_{y}^{2}+m^{2}}{\mu
B^{\prime}}}\left(  se^{-i\frac{\pi}{4}}\sqrt{\mu B^{\prime}}\left(
x-\tfrac{2\varsigma p_{0}-\mu B}{\mu B^{\prime}}\right)  \right)  \right\}  
\label{label eq41} \nonumber
\end{eqnarray}%
\begin{eqnarray}
v_{p_{0},p_{y}}^{\left(  \varsigma,s\right)  }\left(  x,y;t\right)   &
=& \frac{1}{2}\tfrac{e^{-\frac{\pi}{4}\rho}}{\sqrt{\sinh\left(  \pi\frac
{p_{y}^{2}+m^{2}}{\mu B^{\prime}}\right)  }}e^{-ip_{0}t}e^{+ip_{y}%
y}\nonumber\\
&&  \left\{  s\tfrac{\left(  1+s\varsigma\right)  \frac{i\varsigma sp_{y}%
+m}{\sqrt{\mu B^{\prime}}}+\left(  1-s\varsigma\right)  }{2}D_{-\frac
{1+\varsigma s}{2}-i\frac{p_{y}^{2}+m^{2}}{\mu B^{\prime}}}\left(
se^{-i\frac{\pi}{4}}\sqrt{\mu B^{\prime}}\left(  x-\tfrac{2\varsigma p_{0}-\mu
B}{\mu B^{\prime}}\right)  \right)  \right. \\
&&  +\left.  i\tfrac{\left(  1-s\varsigma\right)  \frac{-i\varsigma sp_{y}%
+m}{\sqrt{\mu B^{\prime}}}+\left(  1+s\varsigma\right)  }{2}e^{-i\frac{\pi}%
{4}s\varsigma}D_{-\frac{1-\varsigma s}{2}-i\frac{p_{y}^{2}+m^{2}}{\mu
B^{\prime}}}\left(  se^{-i\frac{\pi}{4}}\sqrt{\mu B^{\prime}}\left(
x-\tfrac{2\varsigma p_{0}-\mu B}{\mu B^{\prime}}\right)  \right)  \right\}
\label{label eq42} \nonumber.
\end{eqnarray}
In matrix form, we can write the eigenspinors for $\varsigma=+1$%
\begin{eqnarray}
&&
\Psi_{p_{0}p_{y}}^{\left(  +1,+1\right)  }\left(  x,y;t\right)  =\frac{1}%
{2}\tfrac{e^{-\frac{\pi}{4}\rho}}{\sqrt{\sinh\left(  \pi\rho\right)  }%
}e^{-ip_{0}t}e^{+ip_{y}y}\left(
\begin{array}
[c]{c}%
\frac{ip_{y}+m}{\sqrt{\mu B^{\prime}}}D_{-1-i\rho}\left(  
z_+\right)  -ie^{-i\frac{\pi}{4}}D_{-i\rho}\left(  z_+\right)
\\
\frac{ip_{y}+m}{\sqrt{\mu B^{\prime}}}D_{-1-i\rho}\left(  z_+\right)  +ie^{-i\frac{\pi}{4}}D_{-i\rho}\left(  sz_+\right)
\end{array}
\right)  \label{label eq43}\\
&&
\Psi_{p_{0}p_{y}}^{\left(  +1,-1\right)  }\left(  x,y;t\right)  =\frac{1}%
{2}\tfrac{e^{-\frac{\pi}{4}\rho}}{\sqrt{\sinh\left(  \pi\rho\right)  }%
}e^{-ip_{0}t}e^{+ip_{y}y}\left(
\begin{array}
[c]{c}%
-D_{-i\rho}\left(  -z_+\right)  +i\frac{ip_{y}+m}{\sqrt{\mu
B^{\prime}}}e^{i\frac{\pi}{4}}D_{-1-i\rho}\left(  -z_+\right)
\\
D_{-i\rho}\left(  -z_+\right)  +i\frac{ip_{y}+m}{\sqrt{\mu
B^{\prime}}}e^{i\frac{\pi}{4}}D_{-\frac{1-\varsigma s}{2}-i\rho}\left(
-z_+\right)
\end{array}
\right)  \label{label eq44}%
\end{eqnarray}
and for $\varsigma=-1$%
\begin{eqnarray}
 &&
\Psi_{p_{0}p_{y}}^{\left(  -1,+1\right)  }\left(  x,y;t\right)  =\frac{1}%
{2}\tfrac{e^{-\frac{\pi}{4}\rho}}{\sqrt{\sinh\left(  \pi\rho\right)  }%
}e^{-ip_{0}t}e^{+ip_{y}y}\left(
\begin{array}
[c]{c}%
D_{-i\rho}\left(  z_-\right)  -i\frac{ip_{y}+m}{\sqrt{\mu
B^{\prime}}}e^{i\frac{\pi}{4}}D_{-1-i\rho}\left(  z_-\right)
\\
D_{-i\rho}\left(  z_-\right)  +i\frac{ip_{y}+m}{\sqrt{\mu
B^{\prime}}}e^{i\frac{\pi}{4}}D_{-1-i\rho}\left(  z_-\right)
\end{array}
\right)  \label{label eq45}\\
&&
\Psi_{p_{0}p_{y}}^{\left(  -1,-1\right)  }\left(  x,y;t\right)  =\frac{1}%
{2}\tfrac{e^{-\frac{\pi}{4}\rho}}{\sqrt{\sinh\left(  \pi\rho\right)  }%
}e^{-ip_{0}t}e^{+ip_{y}y}\left(
\begin{array}
[c]{c}%
-\frac{ip_{y}+m}{\sqrt{\mu B^{\prime}}}D_{-1-i\rho}\left(  -z_-\right)  +ie^{-i\frac{\pi}{4}}D_{-i\rho}\left(  -z_-\right) \\
\frac{ip_{y}+m}{\sqrt{\mu B^{\prime}}}D_{-1-i\rho}\left(  -z_-\right)  +ie^{-i\frac{\pi}{4}}D_{-i\rho}\left(  -z_-\right)
\end{array}
\right)  \label{label eq46}%
\end{eqnarray}
with the change $z_\varsigma=e^{-i\frac{\pi}{4}}\sqrt{\mu B^{\prime}}\left(  x-\tfrac{2\varsigma p_{0}-\mu B}{\mu
B^{\prime}}\right)  $ and $\rho=\frac{p_{y}^{2}+m^{2}}{\mu B^{\prime}}$. 
These are exactly the  wanted eigenspinors corresponding to the continuum energy spectrum $E=p_0$, %our system,
which can be used to deal with different properties of the graphene systems 
\cite{Novoselov} and related matters.
%Thus the
%wave functions related to our problem have exactly been calculated, 
%which remains among the most interesting results derived right now.

Now let us examine the asymptotic behavior of the obtained eigenspinors. Indeed, recalling the 
limiting cases when $\ z\longrightarrow \infty $ \cite{Abramowitz}  
\begin{eqnarray}
D_{\protect\nu }\left( z\right)  &\longrightarrow &z^{\protect\nu }e^{-\frac{%
1}{4}z^{2}}, \qquad \left\vert \arg z\right\vert \leq \frac{\protect\pi }{2} \\
D_{\protect\nu }\left( z\right)  &\longrightarrow &z^{\protect\nu }e^{-\frac{%
1}{4}z^{2}}+\frac{\protect\sqrt{2\protect\pi }}{\Gamma \left( -\protect\nu %
\right) }\left( -z\right) ^{-\protect\nu -1}e^{\frac{1}{4}z^{2}}, \qquad 
\frac{\protect\pi }{2}\leq\left\vert \arg z\right\vert \leq \pi, \left\vert \arg (-z)\right\vert \leq \frac{\pi}{2}.
\protect\end{eqnarray}
These results can be used to show that the obtained eigenspinors reduce to
the following two components for $\varsigma =+1$ and $x\longrightarrow +\infty $
\begin{eqnarray}
\Psi _{p_{0},p_{y}}^{\left( +1,+1\right) }\left( x,y;t\right)
&\longrightarrow& \frac{1}{2}\tfrac{e^{-\frac{\pi }{2}\rho }}{\sqrt{\sinh
\left( \pi \rho \right) }}e^{-ip_{0}t}e^{+ip_{y}y}\left( \mu B^{\prime
}\right) ^{-1-i\frac{\rho }{2}} 
\left( x-\tfrac{2\varsigma p_{0}-\mu B}{\mu B^{\prime }}\right) ^{-1-i\rho} \\
&&
e^{\frac{i}{4}\mu B^{\prime }\left( x-\tfrac{2p_{0}-\mu B}{\mu B^{\prime }}%
\right) ^{2}}e^{i\frac{\pi }{4}}\left( 
\begin{array}{c}
ip_{y}+m-\mu B^{\prime }x+2p_{0}-\mu B \\ 
ip_{y}+m+\mu B^{\prime }x-2p_{0}+\mu B%
\end{array}%
\right)  \nonumber\\
 \Psi_{p_{0},p_{y}}^{\left(  +1,-1\right)  }\left(  x,y;t\right)
&\longrightarrow& \frac{1}{2}\tfrac{e^{\frac{\pi}{2}\rho}}{\sqrt{\sinh\left(
\pi\rho\right)  }}e^{-ip_{0}t}e^{+ip_{y}y}\left(  \mu B^{\prime}\right)
^{-1-i\frac{\rho}{2}}\left(  x-\tfrac{2p_{0}-\mu B}{\mu B^{\prime}}\right)
^{-1-i\rho}\\
&&  e^{\frac{i}{4}\mu B^{\prime}\left(  x-\tfrac{2p_{0}-\mu B}{\mu
B^{\prime}}\right)  ^{2}}\left(
\begin{array}
[c]{c}%
-\mu B^{\prime}x+2p_{0}-\mu B+ip_{y}+m\\
\mu B^{\prime}x-2p_{0}+\mu B+ip_{y}+m
\end{array}
\right)  \nonumber\\
&& +\frac{1}{2}\tfrac{\sqrt{2\pi}e^{\frac{\pi}{2}\rho}}{\sqrt{\sinh\left(
\pi\rho\right)  }}e^{-ip_{0}t}e^{+ip_{y}y}\left(  \mu B^{\prime}\right)
^{-\frac{1}{2}+i\frac{\rho}{2}}\left(  x-\tfrac{2p_{0}-\mu B}{\mu B^{\prime}%
}\right)  ^{i\rho-1}\nonumber\\
&& e^{-\frac{i}{4}\mu B^{\prime}\left(  x-\tfrac{2p_{0}-\mu B}{\mu
B^{\prime}}\right)  ^{2}}e^{i\frac{\pi}{4}}\left(
\begin{array}
[c]{c}%
-\frac{1}{\Gamma\left(  i\rho\right)  }+i\frac{1}{\Gamma\left(  1+i\rho
\right)  }\left(  ip_{y}+m\right)  \left(  x-\tfrac{2p_{0}-\mu B}{\mu
B^{\prime}}\right)  \\
\frac{1}{\Gamma\left(  i\rho\right)  }+i\frac{1}{\Gamma\left(  1+i\rho\right)
}\left(  ip_{y}+m\right)  \left(  x-\tfrac{2p_{0}-\mu B}{\mu B^{\prime}%
}\right)
\end{array}
\right)\nonumber
\end{eqnarray}
and the remaining ones for $x\longrightarrow -\infty $ and  $\varsigma =-1$ for both limits can be 
worked out in the same manner. These summarize the most
interesting results derived so far.

%%%%%%%%%%%%%%%%%%%%%%%%%%%%%%%%%%%%%%%%%%%%%%%%%%%%%%%%%%%%%%%%%%%%%%%%
\section{The probability of pair creation in vacuum in 2+1 dimensions}
%%%%%%%%%%%%%%%%%%%%%%%%%%%%%%%%%%%%%%%%%%%%%%%%%%%%%%%%%%%%%%%%%%%%%%%%%%

Recalling that the {production rate of neutral fermions through the Pauli interaction in 3+1
dimensional was  studied in\cite{Hyun}.}
{Let us look for the probability of pair creation in vacuum in (2+1)-dimensions 
corresponding to our system. Indeed, we start with the relation
gives the effective action} \cite{Lin}
\begin{equation}
\ln S_{\text{eff}}\left[  x,y,t\right]  =\Tr\ln\left[  \left(  \gamma^{\mu
}p_{\mu}-\frac{\mu}{2}\sigma^{\mu\nu}F_{\mu\nu}-m\right)  \frac{1}{\gamma
^{\mu}p_{\mu}-m}\right]  .\label{eq47}%
\end{equation}
Using the charge conjugation matrix $C$ \cite{Zuber}%
\begin{equation}
C\gamma_{\mu}C^{-1}=-\gamma_{\mu}^{T},\qquad C\sigma^{\mu\nu}C^{-1}%
=-\sigma^{T\mu\nu}\label{eq48}%
\end{equation}
to write \eqref{eq47} as
\begin{equation}
2\ln S_{\text{eff}}\left[  x,y,t\right]  =\Tr\ln\left[  \left(  \left(
\gamma^{\mu}p_{\mu}-\frac{\mu}{2}\sigma^{\mu\nu}F_{\mu\nu}\right)  ^{2}%
-m^{2}\right)  \frac{1}{p^{2}-m^{2}}\right]  \label{eq49}.
\end{equation}
By using the identity%
\begin{equation}
\ln\frac{a}{b}=\int_{0}^{+\infty}\frac{d\lambda}{\lambda}\left(
e^{i\frac{\lambda}{2}\left(  b+i\epsilon\right)  }-e^{i\frac{\lambda}%
{2}\left(  a+i\epsilon\right)  }\right)  \label{eq50}%
\end{equation}
we show that the probability, per unit time and per unit area (i.e. $\int
 dtdxdy=1)$, of neutral particle-antiparticle pair creation in the
vacuum takes the form
\begin{equation}
\mathcal{W}=\operatorname{Re}\int_{0}^{\infty}\frac{d\lambda}{\lambda
}\Tr\left[  e^{\left(  i\frac{\lambda}{2}\left(  \mathcal{H}+i\varepsilon
\right)  \right)  }-e^{\frac{i\lambda}{2}\left(  p_{0}^{2}-p_{x}^{2}-p_{y}%
^{2}-m^{2}+i\varepsilon\right)  }\right]  \label{eq51}.
\end{equation}
{Replacing the Hamiltonian} $\cal{H}$ \eqref{eq14} to
obtain
\begin{equation}
\mathcal{W}=2\operatorname{Re}\int_{0}^{\infty}\frac{d\lambda}{\lambda
}e^{-\frac{i\lambda}{2}m^{2}}\Tr\left[  \int_{-\infty}^{+\infty}\frac{dp_{0}%
}{2\pi}
%\int_{-\infty}^{+\infty}
\frac{dp_{y}}{2\pi}e^{-\frac{i\lambda}{2}%
p_{y}^{2}}K^{os}\left(  x,x;\lambda\right)  \cosh\left(  \tfrac{\lambda\mu
B^{\prime}}{4}\right)  -\frac{\pi}{\left(  2\pi\right)  ^{3}\frac{\lambda}{2}%
}\sqrt{\frac{\pi}{\frac{i\lambda}{2}}}\right]  \label{eq52}%
\end{equation}
where%
\begin{equation}
K^{os}\left(  x,x;\lambda\right)  =\sqrt{\tfrac{\tfrac{\mu B^{\prime}}{2}%
}{2i\pi\sinh\left(  \tfrac{\lambda\mu B^{\prime}}{2}\right)  }}\exp\left(
i\frac{2}{\mu B^{\prime}}\left(  p_{0}-\varsigma\mu\frac{B+B^{\prime}x}%
{2}\right)  ^{2}\tanh\left(  \tfrac{\lambda\mu B^{\prime}}{4}\right)  \right)
\label{eq53}.
\end{equation}
Then by performing the integration over $p_{0}$ and $p_{y}$ to obtain
\begin{equation}
\sqrt{\tfrac{\tfrac{\mu B^{\prime}}{2}}{2i\pi\sinh\left(  \tfrac{\lambda\mu
B^{\prime}}{2}\right)  }}\int_{-\infty}^{+\infty}\frac{dp_{0}}{2\pi}%
e^{i\frac{2}{\mu B^{\prime}}\left(  p_{0}-\varsigma\mu\frac{B+B^{\prime}x}%
{2}\right)  ^{2}\tanh\left(  \tfrac{\lambda\mu B^{\prime}}{4}\right)  }%
=\frac{1}{2\pi}\tfrac{\tfrac{\mu B^{\prime}}{2}}{2\sinh\left(  \tfrac
{\lambda\mu B^{\prime}}{4}\right)  }\label{eq54}%
\end{equation}
and%
\begin{equation}
\int_{-\infty}^{+\infty}\frac{dp_{y}}{2\pi}e^{-\frac{i\lambda}{2}p_{y}^{2}%
}=\frac{1}{2\pi}\sqrt{\frac{\pi}{\frac{i\lambda}{2}}}.\label{eq55}%
\end{equation}
Combining all to write
\begin{equation}
\mathcal{W}=2\operatorname{Re}\int_{0}^{\infty}\frac{d\lambda}{\lambda
}e^{-\frac{i\lambda}{2}\left(  m^{2}-i\varepsilon\right)  }\left(  \frac{\mu
B^{\prime}}{4\left(  2\pi\right)  ^{2}}\sqrt{\frac{\pi}{\frac{i\lambda}{2}}%
}\coth\left(  \tfrac{\lambda\varsigma\mu B^{\prime}}{4}\right)  -\frac{\pi
}{\left(  2\pi\right)  ^{3}\frac{\lambda}{2}}\sqrt{\frac{\pi}{\frac{i\lambda
}{2}}}\right)  \label{eq56}%
\end{equation}
or equivalently
\begin{eqnarray}
\mathcal{W} &  =&\frac{1}{8\pi^{2}}\sqrt{\frac{2\pi}{i}}\int_{0}^{\infty}%
\frac{d\lambda}{\lambda^{\frac{3}{2}}}e^{-\frac{i\lambda}{2}\left(
m^{2}-i\varepsilon\right)  }\left(  \frac{\mu B^{\prime}}{4}\coth\left(
\tfrac{\lambda\mu B^{\prime}}{4}\right)  -\frac{1}{\lambda}\right)
\nonumber\\
&&  +\frac{1}{8\pi^{2}}\sqrt{\frac{2\pi}{-i}}\int_{0}^{\infty}\frac{d\lambda
}{\lambda^{\frac{3}{2}}}e^{\frac{i\lambda}{2}\left(  m^{2}+i\varepsilon
\right)  }\left(  \frac{\mu B^{\prime}}{4}\coth\left(  \tfrac{\lambda\mu
B^{\prime}}{4}\right)  -\frac{1}{\lambda}\right)  .\label{eq57}%
\end{eqnarray}
By making the change of variable $\lambda\longrightarrow\lambda^{\prime
}=e^{i\pi}\lambda$ in the second line of the above expression and after some
steps we get%
\begin{equation}
\mathcal{W}=\frac{1}{8\pi^{2}}\sqrt{\frac{2\pi}{i}}\int_{-\infty}^{\infty
}\frac{d\lambda}{\lambda^{\frac{3}{2}}}e^{-\frac{i\lambda}{2}\left(
m^{2}+i\varepsilon\right)  }\left(  \frac{\mu B^{\prime}}{4}\coth\left(
\tfrac{\lambda\mu B^{\prime}}{4}\right)  -\frac{1}{\lambda}\right)
.\label{eq58}%
\end{equation}
Using the residue theorem for the simple poles  $\lambda=\pm i\frac{4n\pi}{\mu
B^{\prime}}$, to obtain
\begin{equation}
\mathcal{W}=\frac{1}{4\sqrt{\pi}}\sum_{n=1}^{+\infty}\frac{1}{\left(
\frac{4\pi n}{\mu B^{\prime}}\right)  ^{\frac{3}{2}}}e^{-\frac{2n\pi m^{2}%
}{\mu B^{\prime}}}.\label{eq59}
\end{equation}
At this level we have some comments in order. Firstly,
\eqref{eq59} is similar to the result obtained in dealing with the case of an electric field
\cite{Qiong}. Secondly, we  observe that the neutral particle creation probability is
an exponentially decreasing function with respect to the inverse of the $\mu
B^{\prime}$. This result is similar to the Schwinger process of electron-positron pair
creation in the strong electric field\cite{Schwinger}.

%%%%%%%%%%%%%%%%%%%%%%%%%%%%%%%%%%%%%%%%%
\section{Conclusion}
%%%%%%%%%%%%%%%%%%%%%%%%%%%%%%%%%%%%%%%%%

%In the present paper,
We have considered %the solutions of the nergy spectrum of 
%we gave an analytical and exact solution for a 
(2+1)-dimensional Dirac equation related to a relativistic half spin particle in
interaction with inhomogeneous magnetic field. Using the global projection
as well as the path integral formalism to write down the corresponding
propagator as causal Green function in terms of different physical quantities.
These allowed to end up with an interesting relation between the matrix elements
of both projections $S^{c}(\mathbf{x}_{b},\mathbf{x}_{a})$
and $S^{c}_g(\mathbf{x}_{b},\mathbf{x}_{a})$ \eqref{eq10}.
To go further, we have fixed the gauge field in terms of two different
magnetic fields and therefore explicitly got the corresponding 
Hamiltonian. This is used together with the boundary conditions  and the propagator associated to the harmonic oscillator
to map the appropriate Green function in terms of the parabolic cylindrical functions.

Later on, we have used the differential equation of the Green function together with
some relevant properties of the parabolic cylindrical functions to further simply
the matrix elements of the Pauli--Dirac operator.
We have made several algebras based on the residue calculus to finally end up with 
the eigenspinors  in terms  of the physical parameters.  Their asymptotic behaviors
were also examined for the limiting case $x\lga +\infty$ and the value $\varsigma=+1$.
Subsequently, we have used the effective action corresponding to our system to study the neutral particle creation probability. 
Indeed, this allowed to obtain a result depending on the included field $B'$
and the additional spin magnetic moment $\mu$.

%The wave functions have been deduced.

\section*{Acknowledgments}

The authors acknowledge the financial support from King Faisal University. The
present work was done under Project Number 160104 , `Path integral for
Pauli-Dirac particle with anomalous magnetic moment'.

\end{document}